\documentclass{article}

\usepackage{PRIMEarxiv}

\usepackage[utf8]{inputenc} 
\usepackage[T1]{fontenc}    
\usepackage{hyperref}       
\usepackage{url}            
\usepackage{booktabs}       
\usepackage{amsfonts}       
\usepackage{nicefrac}       
\usepackage{microtype}      
\usepackage{lipsum}
\usepackage{fancyhdr}       
\usepackage{graphicx}       
\graphicspath{{media/}}     
\usepackage{svg}
\usepackage{amsmath}
\usepackage{caption}
\title{An end-to-end deep learning pipeline to derive blood input with partial volume corrections for automated parametric brain PET mapping}

\author{
  Rugved Chavan* \\
  Department of Computer Science and Engineering \\
  Department of Radiology and Medical Imaging  \\
  University of Virginia \\
  \texttt{qxk6fb@virginia.edu} \\
  \And
  Gabriel Hyman* \\
  Department of Biomedical Engineering \\
  Department of Computer Science\\
  Department of Radiology and Medical Imaging\\
  University of Virginia\\
  \And
  Zoraiz Qureshi* \\
  Department of Computer Science and Engineering \\
  Department of Radiology and Medical Imaging  \\
  University of Virginia \\
  \And
  Nivetha Jayakumar* \\
  Department of Computer Science and Engineering \\
  Department of Radiology and Medical Imaging  \\
  University of Virginia \\
  \And
  William Terrell \\
  Department of Computer Science and Engineering \\
  Department of Radiology and Medical Imaging  \\
  University of Virginia \\
  \And
  Stuart Berr \\
  Department of Radiology and Medical Imaging \\
  University of Virginia\\
  \And
  David Schiff \\
  Department of Neurology \\
  University of Virginia\\
  \And
  Megan Wardius \\
  Brain Institute \\
  University of Virginia\\
  \And
  Nathan Fountain \\
  Department of Neurology \\
  University of Virginia\\
  \And
  Thomas Muttikkal \\
  Department of Radiology and Medical Imaging \\
  University of Virginia\\
  \And
  Mark Quigg \\
  Department of Neurology \\
  University of Virginia\\
  \And
  Miaomiao Zhang \\
  Department of Computer Science and Engineering \\
  University of Virginia \\
  \And
  Bijoy Kundu \\
  Department of Radiology and Medical Imaging\\
  Department of Biomedical Engineering \\
  University of Virginia \\
  \texttt{bkk5a@virginia.edu}
}

\begin{document}

\maketitle

* Indicate authors contributed equally to this work.

\begin{abstract}
Dynamic 2-[18F] fluoro-2-deoxy-D-glucose positron emission tomography (dFDG-PET) for human brain imaging has considerable clinical potential, yet its utilization remains limited. A key challenge in the quantitative analysis of dFDG-PET is characterizing a patient-specific blood input function, traditionally reliant on invasive arterial blood sampling. This research introduces a novel approach employing non-invasive deep learning model-based computations from the internal carotid arteries (ICA) with partial volume (PV) corrections, thereby eliminating the need for invasive arterial sampling. We present an end-to-end pipeline incorporating a 3D U-Net based ICA-net for ICA segmentation, alongside a Recurrent Neural Network (RNN) based MCIF-net for the derivation of a model-corrected blood input function (MCIF) with PV corrections. The developed 3D U-Net and RNN was trained and validated using a 5-fold cross-validation approach on 50 human brain FDG PET datasets. The ICA-net achieved an average Dice score of 82.18\% and an Intersection over Union of 68.54\% across all tested scans. Furthermore, the MCIF-net exhibited a minimal root mean squared error of 0.0052. The application of this pipeline to ground truth data for dFDG-PET brain scans resulted in the precise localization of seizure onset regions, which contributed to a successful clinical outcome, with the patient achieving a seizure-free state after treatment. These results underscore the efficacy of the ICA-net and MCIF-net deep learning pipeline in learning the ICA structure's distribution and automating MCIF computation with PV corrections. This advancement marks a significant leap in non-invasive neuroimaging.

\end{abstract}


\section{Introduction}
Positron emission tomography (PET) is a medical imaging modality that has been used over three decades \cite{phelps1975application}. Unlike other medical imaging modalities, including magnetic resonance imaging (MRI) and computed tomography (CT)---which provide high-resolution information about morphological and anatomical structures---PET reveals \textit{in vivo} physiological and biochemical function of target organs \cite{muzi2012quantitative}. The fundamental element that facilitates this scintigraphic scan is the radiotracer, a molecule known to target the biological pathway of interest linked to a radioisotope. Upon intravenous injection, the radiotracer is transported and uptaken by tissue of interest. The most commonly used radiotracer is 2-[18F] fluoro-2-deoxy-D-glucose (FDG), which is an analog of glucose and, through its use in PET, allows for \textit{in vivo} visualization of glucose metabolism. With FDG-PET, clinicians can identify pathophysiological regions characterized by abnormal glucose regulation, ranging from states of infection \cite{love2005fdg} to diseases like rheumatoid arthritis \cite{zhou2020fdg} and cancer \cite{muzi2012quantitative}, \cite{grkovski2017monitoring}. In particular, FDG-PET of the human brain has become considerably popular in the recent years and can improve neuropathological evaluation of a variety of brain disease states, including glioblastoma \cite{binneboese2021correlation} \cite{hossain2023multimodal}\cite{schetlick2021parametric}, dementias \cite{brown2014brain},  \cite{landau2011associations}, medically intractable epilepsy \cite{tang2018cpne1}, \cite{seshadri2021dynamic} \cite{quigg2022dynamic}, traumatic brain injury \cite{byrnes2014fdg}, and most recently, COVID-19 \cite{guedj2021fdg}.

\textbf{sFDG-PET vs. dFDG-PET :}
Static FDG-PET (sFDG-PET) is the most conventionally used form of PET for the human brain \cite{seshadri2021dynamic}, \cite{dimitrakopoulou2012quantitative}. The patient is injected with the radiotracer and, typically after a 45-minute window of waiting, is imaged in the scanner for 15 minutes \cite{seshadri2021dynamic}. The result of this process is a single snapshot (averaged over the 15-minute scan) of neuronal glucose metabolism, which is generally only qualitatively interpretable \cite{muzi2012quantitative}. Dynamic FDG-PET (dFDG-PET) of the human brain is a newer and emerging form of PET that allows for both the visualization and quantification of the same biological pathway information, but instead as a function of time, rather than a single snapshot. In dFDG-PET the patient is first put on the scanner and a 60-90 minute scan initiated followed by a slow injection of the radiotracer over a period of 5-10 seconds, thereby capturing data from the point of injection. Glucose metabolic activity of the brain is captured at discrete time points throughout the scan, which results in a 4D-object consisting of multiple 3D-volumes of pathway activity \cite{seshadri2021dynamic}. Recent studies have highlighted the additional clinical utility of dFDG-PET compared to sFDG-PET across a variety of disease states \cite{grkovski2017monitoring}, \cite{tang2018cpne1}, \cite{seshadri2021dynamic}, \cite{dimitrakopoulou2012quantitative}– \cite{wangerin2017virtual}. By capturing both the temporal and spatial pattern of radiotracer uptake, dFDG-PET provides clinicians with substantially more information about \textit{\textbf{in vivo}} pathway activity and even reduces the negative impact of undesired artifacts present in sFDG-PET \cite{muzi2012quantitative}.

\textbf{IDIF:}
Quantitative analysis of these dynamic scans requires an understanding of patient-specific radiotracer biochemistry, which is given by the blood input function. The blood input function 
is a time-activity curve (TAC) that describes the amount of radiotracer in the blood available to tissue as a function of time throughout the course of the scan. The gold standard for measuring this function is through time-distributed arterial blood sampling during the scan, which is costly \cite{vaquero2015positron} and risks patient infection or arterial occlusion \cite{oertel2012necessity}. Researchers, however, have recently discovered a way by which this kinetic radiotracer information can be successfully extracted directly from the image data \cite{wangerin2017virtual}. This non-invasive measurement is known as an image-derived blood input function (IDIF) and monitors how the concentration of radiotracer (which, in dFDG-PET, is highly correlated to voxel intensity) varies across a region-of-interest (ROI) during the scan. The carotid arteries, which supply the brain with oxygenated blood and, thus, radiotracer, are the gold standard ROI to segment out of the scan and from which to derive the IDIF \cite{seshadri2021dynamic}, \cite{croteau2010image}–\cite{lyoo2014image}. 
Despite the benefit of using an IDIF in the place of arterial blood sampling to quantify radiotracer kinetics, the associated annotation process serves as a major roadblock to the wider adoption of dFDG-PET. Manually segmenting out the carotid arteries is a highly imprecise and time-consuming process and requires a trained technician to perform \cite{lyoo2014image}. Recognizing these limitations, one of the two objectives of this work is to implement supervised machine learning (ML) techniques to develop a pipeline capable of automatically segmenting carotid ROIs
and performing subsequent IDIF calculations, ideally facilitating broader adoption of dFDG-PET for human brain imaging.

\textbf{MCIF:}
However, PET scanners typically have inherently low resolution [33], which makes raw IDIFs vulnerable to spillover (SP) from adjacent tissues and partial volume (PV) effects.
Spillover (SP) effects occur when the signal from a high-activity region 'spills over' into adjacent lower-activity regions, leading to overestimation of tracer concentration in these adjacent areas. This phenomenon is particularly evident in small structures adjacent to high uptake tissues. On the other hand, partial volume (PV) effects arise due to the finite spatial resolution of PET scanners, where the measured signal is a blend of activities from different tissues within the same voxel. This blending leads to an underestimation of peak tracer concentration in small structures, as the signal from these structures is diluted by surrounding tissue with lower tracer uptake. 
So these issues arise when analyzing image regions with radii smaller than the spatial resolution of the PET scanner, such as the carotid arteries. To more accurately estimate the blood input function, we correct the IDIF with a 3-compartment model to obtain the Model Corrected Blood Input Function (MCIF). 
In our previous research, we determined the MCIF by regressing the 3-compartment model to the IDIF and the TAC of the tissue surrounding the ICA, utilizing a custom loss function. Due to the large number of parameters of the 3-compartment model (15) and the non-convexity of this optimization task, prior bounds for the parameters had to be manually determined. The second objective of this work is to advance this approach by implementing a deep learning-based calculation of this model. We developed a recurrent neural network (RNN) based architecture, MCIF-net, designed specifically for predicting MCIF from IDIF.

\section{Materials and Methods}
In this section, we will discuss the dataset's collection and preparation, the comprehensive end-to-end pipeline, the architecture of the model, and the evaluation metrics employed for assessing performance. 

\subsection{Dataset}
The dFDG-PET imaging was conducted on a cohort comprising 
of 50 participants. This procedure was performed using the Siemens Biograph time-of-flight mCT scanner. The scans featured a resolution of 400 pixels x 400 pixels x 110 slices x 38 timeframes, and were conducted with time-dependent attenuation correction. The dynamic acquisition process involved the initiation of a 60-minute scan followed immediately with an intravenous administration of approximately 10 mCi FDG, injected over a duration of 10 seconds. All data were reconstructed in a list-mode format. Prior to the dFDG-PET procedure, all 50 patients underwent T1-weighted MPRAGE MPI (256 pixels x 256 pixels x 192 slices) using a Siemens 3T scanner for co-registration and anatomic mapping. 

\begin{figure}[ht]
    \centering
    \includegraphics[width=1\linewidth]{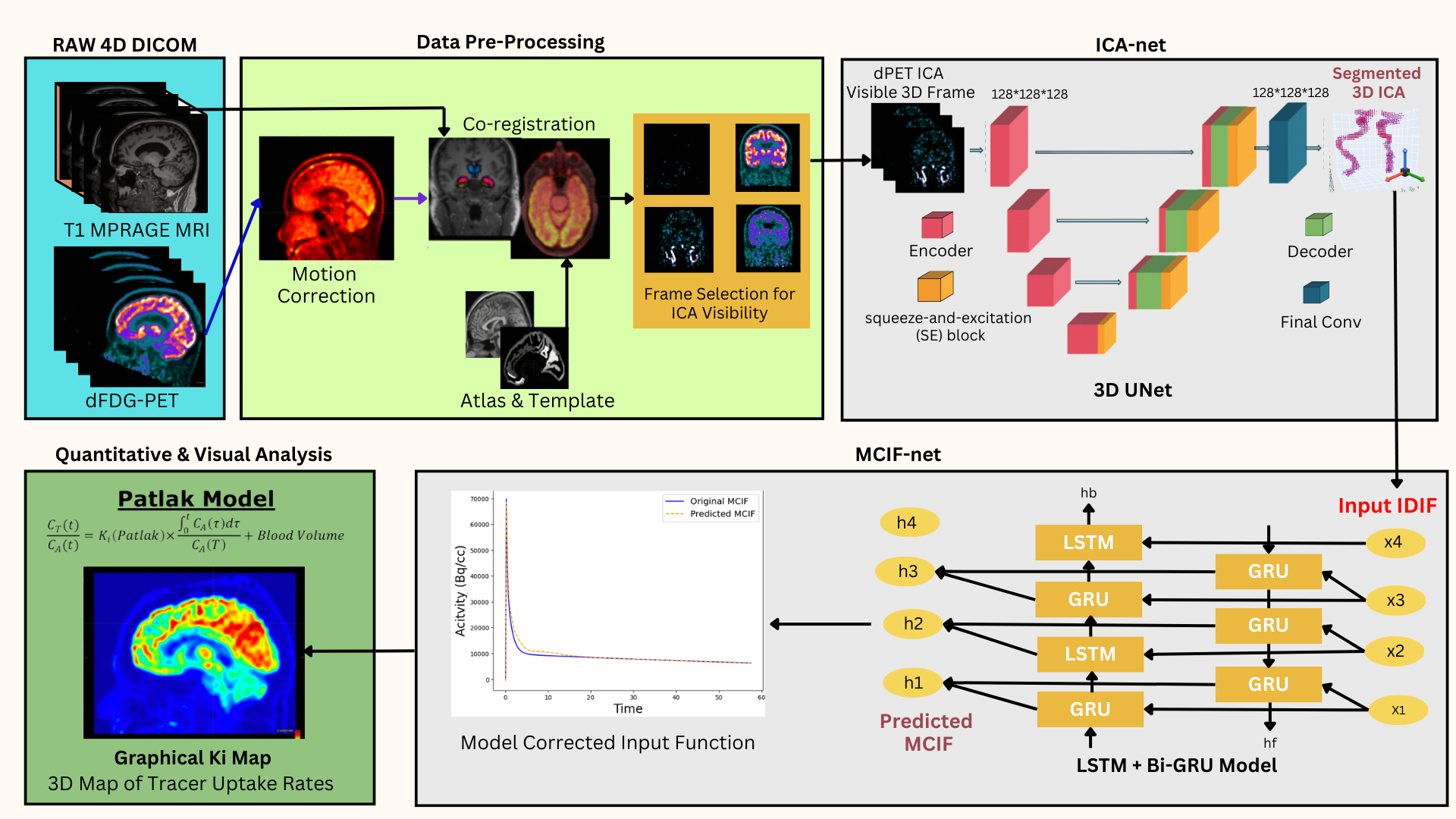}
    \caption{Abstract overview of the complete pipeline with two disjoint supervised models for segmentation and blood input correction}
    \label{absract_fig}
\end{figure}

\subsection{End to End Pipeline}
 The pipeline comprises four distinct phases. \textbf{The first phase} encompasses data preprocessing, which includes motion correction and co-registration of the dynamic FDG-PET (dFDG-PET) data. \textbf{The second phase} utilizes the 4D motion-corrected and co-registered data for internal carotid arteries (ICA) segmentation through ICA-net. \textbf{The third phase} involves the computation of the Image-Derived Blood Input Function (IDIF), monitoring changes in intensity within the ICA region. This phase also includes the application of MCIF-net for correcting PV and SP errors, resulting in the final MCIF. \textbf{The fourth phase} employs the MCIF to calculate the Ki Map using the Patlak model. Subsequently, for patients with epilepsy and known surgical ground truth, we compute the Z-score, normalizing it against the mean and standard deviation (SD) for the whole brain, encompassing 18 super regions per side. \textbf{Figure \ref{absract_fig}} delineates the workflow of this pipeline.

 In this work, we automated an integrated, end-to-end pipeline by developing ICA-net for ICA segmentation and MCIF-net for partial volume corrections. This system is specifically tailored for the automatic identification of seizure foci in human dynamic FDG brain PET imaging.

\subsubsection{Data Pre-processing}
The initial step in pre-processing started with motion correction for the 60-minute acquisition to align and lock the anatomy in the same 3 dimensional space throughout the entire time period. PET data (400 pixels x 400 pixels x 111 slices x 38-time frames) was averaged across the first 14 time frames to create a reference for the proper alignment in the image space. This reference was used to perform a rigid body transform across the 38 frames for motion correction. Next, a new average frame of all the motion corrected PET frames was resliced and co-registered into MRI space using the T1 weighted MRI using non-rigid transforms to generate a transformation matrix. This was used, in turn, to generate a co-registered dynamic PET volume.  All the above registration processes were designed using software from the FMRIB’s Software Library (FSL) tool kit \cite{smith2004advances} \cite{woolrich2009bayesian} \cite{jenkinson2002improved} \cite{jenkinson2012fsl} \cite{jenkinson2001global} \cite{seshadri2021dynamic} (\textbf{Figure 1}).



\subsubsection{Segmentation of Internal Carotid Arteries}

\subsubsubsection{\textbf{Frame-selection:}}



This first step of this algorithm was to perform a center crop of the first frame based on the central position of the carotids with respect to the neck and brain. Next, the algorithm calculated the sum of intensity across all voxels in the cropped region. These two steps were then performed iteratively across the first 10 frames, generating a plot of total intensity as a function of scan timeframe. From this discrete function, the difference between each frame’s summed intensity and the previous frame’s summed intensity was calculated, generating a plot of summed intensity difference with respect to the previous frame as a function of scan timeframe (with the first frame having an intensity difference of its original summed intensity value). With these differences, the algorithm then selected the frame that occurred one frame before the first local maximum. This local maximum indicated a large radiotracer presence distributed throughout the brain, meaning that the majority of the radiotracer had already been transported through the carotids. The previous frame, therefore, represented the frame in which a considerable amount of radiotracer was still being transported through the carotids and the arteries were, thus, most clear for segmentation. 
All incorrectly selected frames were no more than 1 frame off.

\subsubsubsection{\textbf{Semi-automated segmentation:}}

To optimize the efficiency of this research, addressing the time-consuming nature of manual segmentation of carotids, which typically required 35-45 minutes per scan, was essential. Hence, a semi-automated method was devised for generating ground truth annotations across all 50 datasets. This approach utilized the imaging visualization and analysis capabilities of 3D Slicer, particularly leveraging various functions within its Segment Editor module.

Given that the carotid arteries are responsible for supplying blood and, consequently, radiotracer to the brain, they exhibit high-intensity signals in their lumen. Utilizing 3D Slicer’s Threshold function, an intensity threshold was empirically determined—sometimes through trial-and-error—to segment regions of the scan with voxel intensities equal to or greater than this threshold. The ‘Show 3D’ option was then activated to facilitate the visualization of the 3D-rendered segmentation.

Recognizing that the carotids are relatively large and continuous structures within the scanned area, another empirically determined threshold was set using the Islands function. This function, particularly the 'Split islands to segment' option, was employed to isolate structures with a continuous length meeting or exceeding this threshold. Following the execution of these functions, typically only two structures remained. These were amalgamated using the Logical Operator function, applying the ‘Operation: Add to each.’ Subsequently, one of the objects under ‘Name’ was deleted. Manual removal of extraneous regions and noise was performed using the Scissors function.

To ensure accurate annotations, each slice was meticulously reviewed and verified, resulting in carotid segmentations that generally encompassed 40-60 slices. Notably, this methodology consistently produced results in less than 5 minutes. To advance our research, we resolved to utilize the segmented internal carotid arteries (ICA) as the ground truth and the 3D dynamic FDG-PET (dFDG-PET) data, selected by the frame selector algorithm, as the label for our proposed ICA-net model. This model is specifically designed to perform the segmentation task in a matter of seconds, thereby significantly enhancing the efficiency and accuracy of the segmentation process in comparison to Semi-automated methods.

\subsubsection{ICA-net:}

The ICA-net (\textbf{Figure 2}) is a customized model designed for the segmentation of ICA in 3D dFDG-PET scans. The ICA-net model leverages a modified 3D U-Net architecture \cite{cciccek20163d}, specifically adapted for handling the unique challenges presented by dFDG-PET data. The preprocessing of ICA annotations, derived from a semi-automated segmentation method, includes binarization of the volume data. This is achieved through a custom 'binarizeVolume' function, which applies a threshold to convert the scan data into a binary format, thereby simplifying the learning process for the ICA-net segmentation model.

ICA-net's architecture is built on the backbone of VGG16, a well-established model in the field of deep learning. The input to the network is a 3D volume of shape 128x128x128, with a single channel representing grayscale images. The network is trained using a combined loss function consisting of Dice loss and binary cross-entropy (BCE) loss. 

The combined loss function \( L(y_{\text{true}}, y_{\text{pred}}) \) is given by:

\begin{equation}
L(y_{\text{true}}, y_{\text{pred}}) = \text{DiceLoss}(y_{\text{true}}, y_{\text{pred}}) + \text{BCE}(y_{\text{true}}, y_{\text{pred}})
\end{equation}

where Dice loss is defined as:

\begin{equation}
\text{DiceLoss}(y_{\text{true}}, y_{\text{pred}}) = 1 - \text{DiceCoefficient}(y_{\text{true}}, y_{\text{pred}})
\end{equation}

and the Dice coefficient is calculated as:

\begin{equation}
\text{DiceCoefficient}(y_{\text{true}}, y_{\text{pred}}) = \frac{2 \cdot \sum{(y_{\text{true}} \cdot y_{\text{pred}})} + \text{smooth}}{\sum{y_{\text{true}}} + \sum{y_{\text{pred}}} + \text{smooth}}
\end{equation}

with \(\text{smooth}\) being a small constant added to prevent division by zero.

The Combined Dice and BCE Loss function integrates the advantages of both Dice and BCE loss functions to address ICA segmentation challenges:

\begin{enumerate}
  \item \textbf{Class Imbalance Compensation:} Dice Loss is sensitive to the class imbalance due to its formulation:
  \[ \mathcal{L}_{\text{Dice}} = 1 - \frac{2 \sum_{i}^N p_i g_i}{\sum_{i}^N p_i + \sum_{i}^N g_i} \]
  However, BCE compensates by providing a strong gradient for each pixel:
  \[ \mathcal{L}_{\text{BCE}} = -\frac{1}{N} \sum_{i}^N \left[ g_i \log(p_i) + (1 - g_i) \log(1 - p_i) \right] \]

  \item \textbf{Gradient Behavior:} BCE provides a consistent gradient flow for backpropagation, which can help in scenarios where Dice Loss gradients may become sparse, particularly when the overlap between $p_i$ and $g_i$ is minimal.

  \item \textbf{Comprehensive Error Signal:} Combining Dice and BCE Loss functions offers a more holistic error signal for optimization:
  \[ \mathcal{L}_{\text{combined}} = \mathcal{L}_{\text{Dice}} + \mathcal{L}_{\text{BCE}} \]

  \item \textbf{Optimization Landscape:} The combined loss function creates an optimization landscape that leverages the benefits of both loss functions, potentially avoiding the local minima that could be encountered when using Dice Loss alone.

  \item \textbf{Robustness:} The addition of BCE Loss ensures that the model is penalizing individual pixel misclassifications, which can make the segmentation process more robust to noise and outliers.
\end{enumerate}

This combined loss function can therefore facilitate improved model convergence and generalization. Furthermore, the ICA-net employs several advanced techniques to bolster its performance and robustness. Data augmentation plays a pivotal role, wherein each 3D MRI brain image is transformed through randomized rotations (up to ±4 degrees), shifts (up to 5 voxels), and zooms (±0.05 scale factor), subsequently adjusted to a uniform size of 128x128x128 voxels through zero-intensity padding and cropping. This process effectively expands the training dataset, enhancing the model's exposure to diverse clinical scenarios. Additionally, the Adam optimizer is utilized for efficient network training, with its adaptive learning rate capabilities. The model also implements early stopping and model checkpointing strategies to prevent overfitting and to retain the best model weights during training. Finally, the model's parameters, totaling over 71 million, underscore its complexity and capacity to capture detailed features essential for accurate ICA segmentation.

\begin{figure}[ht]
    \centering
    \includegraphics[width=1\linewidth]{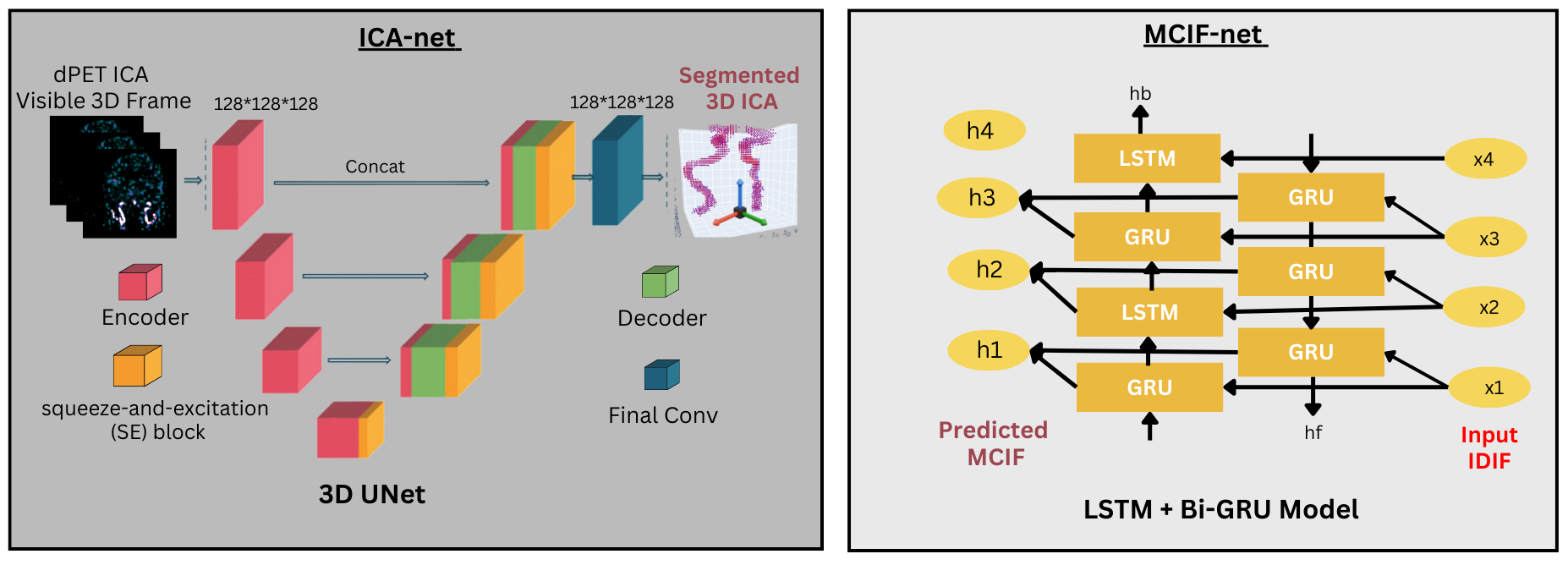}
    \caption{Deep Learning Models}
\end{figure}

\subsubsection{MCIF-net:}
To generate the parametric brain PET maps, MCIF is computed by optimizing the IDIF derived from the ICA as described \cite{massey2021model}, to account for partial volume recovery of the blood input.

The MCIF-net model (\textbf{Figure 2}) is developed using a hybrid recurrent neural network architecture, specifically designed to handle the time-series data inherent in dynamic PET imaging. This architecture integrates both LSTM \cite{hochreiter1997long} and Bi-directional GRU layers \cite{cho2014properties}, leveraging the strengths of each to enhance the processing of PET scan data. The LSTM layers, with increasing complexity from 50 to 200 neurons, are adept at capturing long-term dependencies, while the inclusion of Bi-directional GRU layers provides a richer context by processing data in both forward and backward directions. Dropout layers are interspersed throughout to prevent overfitting. TimeDistributed Dense layers are incorporated to maintain the temporal structure of the output, aligning with the sequential nature of the input.

Data preparation for this combined architecture involved reshaping it into a format suitable for time-series analysis, specifically [samples, time steps, features]. The data was split according 5 fold cross validation.

Compiled with the Adam optimizer and trained on a mean squared error loss function, MCIF-net underwent fine-tuning over 1000 epochs with a batch size of 32. With a total of 5,127,801 trainable parameters, the model's sophisticated architecture is strategically designed to maximize accuracy in predicting MCIF, a critical factor in PET scan analysis. This implementation of a combined LSTM and Bi-directional GRU approach in MCIF-net exemplifies the integration of advanced machine learning techniques in medical imaging, highlighting its capability to provide nuanced and accurate analysis of complex time series data.

\subsection{Evaluation Metrics}

ICA-net: The evaluation of the model's performance is rigorous. We utilize the 5-Fold cross-validation technique, ensuring a thorough and unbiased assessment of the model. Metrics such as the Jaccard score and Dice coefficient are calculated for each fold, providing a comprehensive view of the model's segmentation capabilities. 
The overall metrics, including the mean and standard deviation across all folds, provide a holistic view of the model's performance, demonstrating its efficacy in segmenting the carotid arteries in dFDG-PET scans.

MCIF-net: The evaluation of the MCIF-net, an RNN-based regression model, is centered around two key metrics: Mean Squared Error (MSE) and Mean Absolute Error (MAE). These metrics are particularly suited for regression models as they provide a clear indication of the model's accuracy in predicting continuous values.

\section{Results}

We trained and evaluated the ICA-net and MCIF-net models separately, employing a standard 5-fold cross-validation approach. Furthermore, we validated our complete pipeline against ground truth epilepsy data by calculating z-scores and identifying the seizure onset zone.

\begin{table}[h]
\centering
\caption{ICA-net Fold-wise Metrics with Mean and Standard Deviation}
\label{tab:ICA_net_fold}
\begin{tabular}{lcccc@{}}
\hline
Metric & Train Loss & Val Loss & Val Dice & Val IoU \\
\hline
Fold 1 & 0.0473 & 0.221 & 0.7865 & 0.6529 \\
Fold 2 & 0.0531 & 0.1005 & 0.8602 & 0.7367 \\
Fold 3 & 0.052 & 0.1186 & 0.8575 & 0.7232 \\
Fold 4 & 0.0431 & 0.154 & 0.8466 & 0.7008 \\
Fold 5 & 0.044 & 0.2401 & 0.7586 & 0.6136 \\
\midrule
Mean & 0.047901 & 0.16685 & 0.8218 & 0.685423 \\
Std & 0.00404 & 0.05513 & 0.04147 & 0.04582 \\
\hline
\end{tabular}
\end{table}


\subsection{ICA-net}
\begin{figure}
    \centering
    \includegraphics[width=1\textwidth]{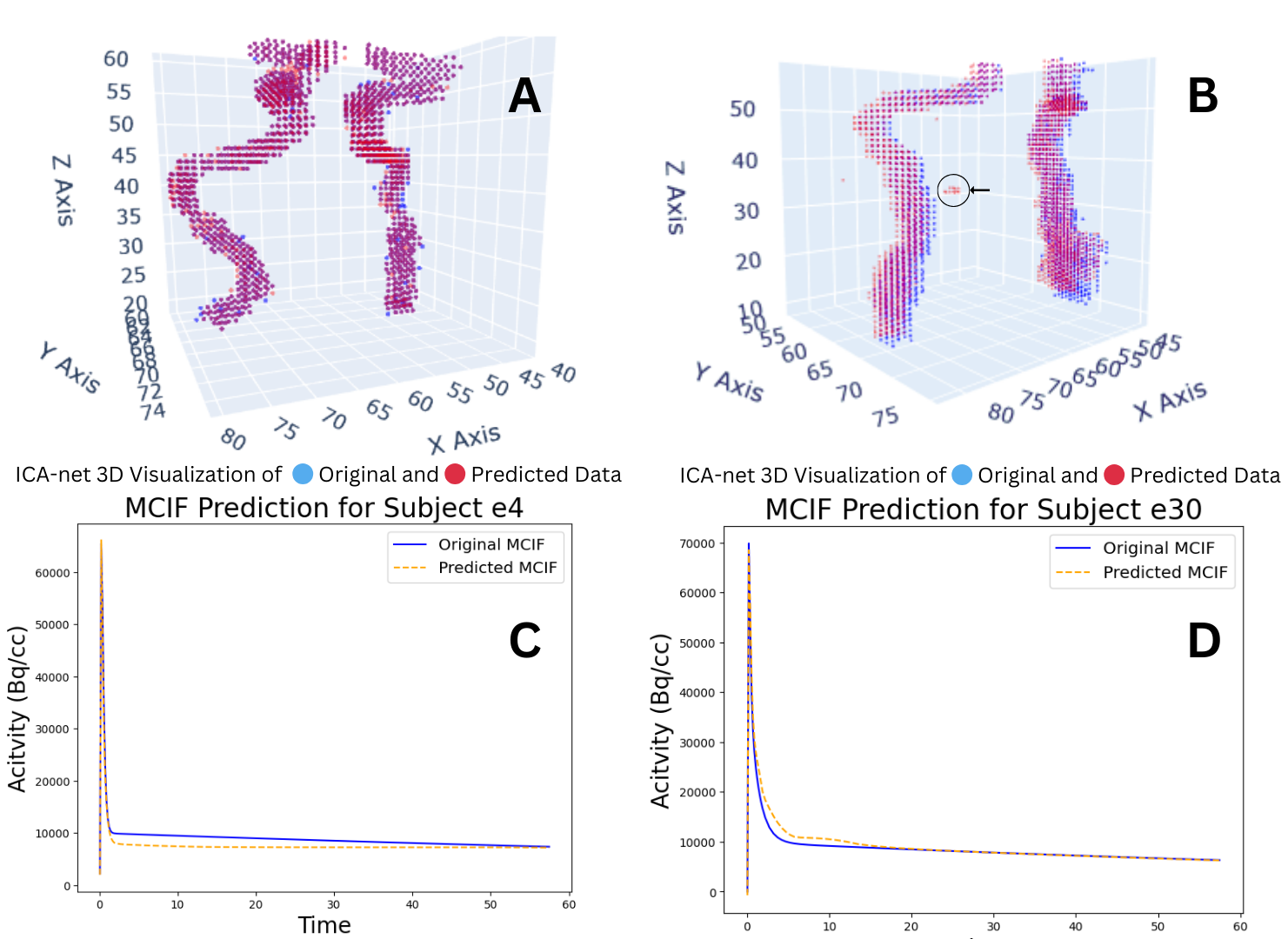}
    \caption{Prediction results of ICA-net and MCIF-net on two test subjects}
    \label{fig:example_image}
\end{figure}

The ICA-net model, developed on the foundation of the UNet architecture, underwent training across 120 epochs, leveraging a 5-fold cross-validation with a leave-one-out approach. This rigorous training regimen yielded a mean Dice coefficient of 82.18\% and a Jaccard score of 68.5423 \% as shown in \textbf{Table} \ref{tab:ICA_net_fold}. \textbf{Figure} \ref{fig:example_image}\textbf{A} illustrates the segmented results, where it is distinctly observable how the predicted ICA closely aligns with the original ICA, demonstrating the model's precision in segmentation. Notably, the analysis reveals that the predicted IDIF points sometimes adds more clusters, slightly distinct from the primary two ICA clusters as shown in \textbf{Fig 3 B} (arrow). Upon exclusion of these outlier clusters, we anticipate an increase in accuracy ranging from approximately 1.5\% to 3.5\%, further enhancing the model's performance.

\begin{table}[ht]
\centering
\caption{Performance of MCIF-net with Different Architectures}
\begin{tabular}{lcc}
\hline
Architecture & MAE & MSE \\
\hline
LSTM & 0.1277 & 0.0991 \\
GRU & 0.1253 & 0.0478 \\
Bi-directional GRU &  0.1183 & 0.0351 \\
\textbf{Comb Bi-GRU + LSTM} & 0.0526 & 0.0052 \\
\hline
\end{tabular}
\label{tab:mcif_net_performance}
\end{table}

\subsection{MCIF-net}
The MCIF-net, initially built on a customized LSTM architecture, was trained for 1000 epochs again using 5 fold cross validation. This initial approach resulted in a Mean Absolute Error (MAE) of 0.1277073 and a Mean Squared Error (MSE) of 0.09911411. We then modified the MCIF-net architecture to incorporate GRU, which achieved an MAE of 0.1253767 and an MSE of 0.04786217. Further adaptations using a Bi-directional GRU resulted in an MAE of 0.1183242 and an MSE of 0.03519516. Additionally, a combination of Bi-directional GRU and LSTM yielded the best results with an MAE of 0.0526 and an MSE of 0.0052, as detailed in
\textbf{Table} \ref{tab:mcif_net_performance} 
. Example predicted MCIF compared to model or original MCIF are shown for 2 example data sets \textbf{Figure 3 C-D}.  

The performance of both the networks (ICA-net and MCIF-net) in this work are an improvement over our previous work due to improved networks and an increase in data sets for model training \cite{jayakumar2023deep}. 

\begin{table}[h]
\centering
\caption{MCIF-net Model Performance Metrics Across Folds}
\label{table:metrics}
\begin{tabular}{@{}lccccc@{}}
\toprule
Metric & Train Loss & Val Loss & Val MSE & Val MAE & Val RMSE \\ 
\midrule
Fold 1       & 0.0094     & 0.0063   & 0.0063  & 0.0640  & 0.0791   \\
Fold 2       & 0.0042     & 0.0035   & 0.0035  & 0.0302  & 0.0594   \\
Fold 3       & 0.0078     & 0.0039   & 0.0039  & 0.0448  & 0.0621   \\
Fold 4       & 0.0129     & 0.0072   & 0.0072  & 0.0707  & 0.0846   \\
Fold 5       & 0.0090     & 0.0050   & 0.0050  & 0.0532  & 0.0710   \\
\midrule
Average      & 0.0087     & 0.0052   & 0.0052  & 0.0526  & 0.0712   \\
\bottomrule
\end{tabular}
\label{tab:mcif_net_cross_f_performance}
\end{table}

\begin{figure}[ht]
    \centering
    \includegraphics[width=1\linewidth]{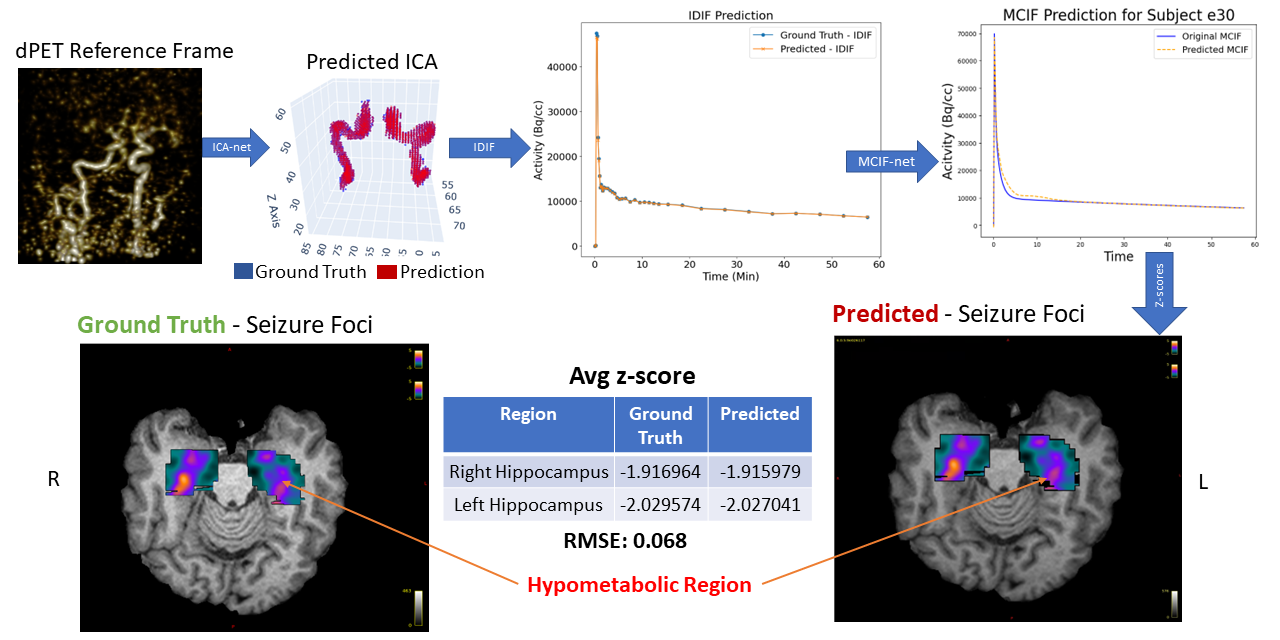}
    \caption{Comprehensive End-to-End Predictive Analysis of Ground Truth Data 'E30'}
    \label{End-to-End-results}
\end{figure}

\subsection{Z-score results}
In this study, we focused on a single epilepsy patient, referred to as 'e30', who had confirmed surgical ground truth. We employed our end-to-end pipeline models, ICA-net and MCIF-net, to analyze this case. The predictive models were utilized to calculate Z-scores for the Ki values, which were normalized against the mean and standard deviation (SD) of the whole brain, encompassing 18 super regions on each side \cite{seshadri2021dynamic}. A Z-score cutoff of less than -2 SD was established to identify hypometabolic regions. This approach successfully highlighted significant hypometabolism in the patient's left hippocampus, with a noteworthy Z-score of -2.027041.

The accuracy of our models was further validated by comparing the Z-score calculations against the ground truth derived from a 15 parameter compartment model. The resultant average root mean square error (RMSE) was 0.068, indicating a high degree of precision in the model's predictions. This precise localization as shown in \textbf{Figure}\ref{End-to-End-results} was critical, as traditional diagnostic methods, including standardized uptake value (SUV) and static PET (sPET), had previously failed to identify the affected area. Our findings were further corroborated by intracranial EEG results, which localized the seizure onset zone (SOZ) to the left hippocampus.

The clinical relevance of our study was underscored by the patient's treatment outcome. Patient 'e30' underwent laser interstitial thermal therapy (LITT) targeting the identified hypometabolic region based on invasive intracranial monitoring in the right hippocampus, which agreed with our model predictions \textbf{(Figure 4)}. This intervention resulted in a significant clinical improvement, with the patient achieving a seizure-free status for a period of 3 months by the follow-up on September 9. This outcome not only validates the effectiveness of our model but also highlights its potential in enhancing clinical decision-making for epilepsy treatment.

\section{Discussion}

These results demonstrate the ability of a 3D-U-Net-based segmentation approach to effectively identify and segment out the carotid arteries in dFDG-PET of the human brain. Additionally, using network-predicted IDIFs to calculate important downstream dFDG-PET metrics, like the Ki value \cite{tang2018cpne1}, \cite{guedj2021fdg}, \textbf{}would serve as more clinically relevant measurements on which to base the evaluation of this model with reference to these same downstream metrics resulting from respective ground truth annotations.

Continuing, testing more datasets would also allow for better evaluation of this model’s generalizability. A dataset of 50 scans is extremely limited and, having already demonstrated successful model learning on so few data, performance would likely improve substantially with the acquisition of a more extensive and variable dataset. Additionally, it would be of interest to assess model performance on scans that made use of different radiotracers. As discussed earlier, radiotracer selection is specific to the biological pathway relevant to the disease state being assessed and there are many important diseases that are not typically studied using dFDG-PET. By both training and testing on datasets that use radiotracers other than FDG, the utility of automatic carotid segmentation and IDIF computation could be extended to the study of a larger scope of neurological disorders, making dPET yet more clinically robust and attractive.

Another way in which this segmentation model could be further optimized is through additional exploration of other training parameters, network architectures and loss functions. Due to the limitations of computational power available for this project, it was not possible to more exhaustively investigate the effect of modifying how the model was trained, like factor of data augmentation (like using a generative adversarial network (GAN) \cite{sandfort2019data} instead of simply applying affine transformations) and batch size, which could potentially lead to the discovery of a more ideal training protocol and effectuate more robust segmentation. These computational limitations also prevented the use of the full reference frame in training, which needed to be cropped, likely leaving out important features from which the model could learn the structure of carotids. Similarly, running a grid-search, for example, on different combinations of architectural hyper-parameters, like depth, base filter and dropout layer rate, could facilitate the discovery of a more effective model.

In the realm of network architectures, while 3D U-Net enjoys popularity, the integration of transformer-based architectures \cite{valanarasu2021medical} (UNetR \cite{hatamizadeh2021unetr} ) could bring substantial benefits to medical imaging segmentation, particularly in dPET. Transformers, known for their proficiency in capturing long-range dependencies and context, could significantly improve the delineation of complex structures in medical images. Their ability to focus on relevant parts of an image makes them particularly suitable for handling the intricacies and variability in dPET images.

Similarly, the potential of sequence-to-sequence architectures \cite{ahmed2022transformers} for our MCIF-net should be considered. Given their success in processing and predicting time-series data, these architectures could accurately model the dynamic tracer distribution in PET scans, which is vital for precise MCIF estimation. This approach could lead to more accurate corrections of SP and PV effects, which are crucial in small vessel structures like carotids.

Additionally, a systematic exploration of architectural hyperparameters through techniques like grid search could lead to the discovery of a more effective model. Adjusting parameters such as network depth, base filter count, and dropout rates could unveil optimal configurations for the unique challenges of dPET image segmentation. Furthermore, considering alternative architectures like feature-pyramid-based models \cite{li2021ifpn}, and statistical-analysis-based techniques like local means analysis (LMA) and soft-decision similar component analysis (SCA) \cite{zanotti2009comparison}, may provide novel avenues for carotid artery segmentation. Finally, the exploration of different loss functions, especially hybrid loss functions that have demonstrated success in relevant segmentation tasks \cite{huang2021learning}, \cite{yang2022mhnet}, could reveal which metrics most closely correlate with low IDIF error resulting from network predictions and guide the selection of the most optimal loss function for this problem.


To successfully allow for fully automatic IDIF computation, the developed frame-selection, as is explained under Frame-selection, must be automated. Frame-selection is crucial as it provides the model with the most ideal frame for selection, making accurate automatic segmentation more likely if effective. Automated algorithms based on intensity analysis, which include intensity and continuous-structure-length thresholding, as well as the use of a trained 3D CNN model, could be explored to enhance the alignment and accuracy of automatic frame-selection. These approaches aim to more closely match the results of an automated algorithm with those of manually determined ideal reference frames. Furthermore, the MCIF-net's evolution through various LSTM, GRU and Fusion architectures culminates in its superior performance in error metrics, signifying its effectiveness in dPET imaging analysis.

\section{Conclusion}
This study's effective training and evaluation of the ICA-net and MCIF-net models represent a significant stride in medical imaging and epilepsy treatment. Both models, validated through a 5-fold cross-validation approach, have shown high accuracy. The ICA-net excelled in segmentation accuracy, and the MCIF-net, through various LSTM and GRU architectures, demonstrated improved performance in error metrics.

Crucially, in a case study of an epilepsy patient, the models accurately identified hypometabolic regions, with the calculated Z-scores aligning with the seizure onset zone. This precision, especially in comparison to traditional diagnostic methods, highlights the models' clinical relevance.

This success underscores the potential of our approach in enhancing clinical decision-making and indicates a promising direction for future research in medical diagnostics and patient-specific treatment strategies.

\section*{Acknowledgments}
This work was supported in part by grants from the University of Virginia Brain Institute (to DS, TM, MQ and BK), Ivy innovation funds (MQ and BK), Commonwealth commercialization funds (CCF) (to BK) from Virginia Innovation Partnership Corporation and start-up funds from the Department of Radiology and Medical Imaging at the University of Virginia (to BK).

\bibliographystyle{unsrt}  
\bibliography{references}

\end{document}